\documentclass[12pt]{article}
\usepackage{amsfonts,amsmath,amssymb}
\usepackage{hyperref}
\usepackage{cite}
\usepackage{epsfig}
\usepackage{paralist}
\usepackage{fancyhdr}
\usepackage{tikz}
\usepackage{tkz-euclide}
\usetikzlibrary{decorations.pathmorphing,patterns,calc,snakes,arrows}

\usepackage{graphicx}
\usepackage{xcolor}
\numberwithin{equation}{section}
\usepackage[vcentermath]{youngtab}
\usepackage{etex}
\usepackage{braket}
\usepackage{float}

\def\spa#1{\phantom{\fbox{\rule[-#1cm]{0cm}{0cm}}}}

\def\be{\begin{equation}}
\def\ee{\end{equation}}
\def\bea{\begin{eqnarray}}
\def\eea{\end{eqnarray}}

\def\del{\partial}
\def\nn{\nonumber}

\renewcommand{\thefootnote}{\fnsymbol{footnote}}

\makeatletter
\g@addto@macro\bfseries{\boldmath}
\makeatother

\def\p{\partial}

\def\Tb{{\bar{T}}}

\begin{document}

\hfuzz=100pt
\title{{\Large \bf{$T\bar{T}$-deformed correlators from a 2D gravity description}}}
\date{}
\author{Shinji Hirano$^{a, b}$\footnote{
	e-mail:
	\href{mailto:shinji.hirano@gmail.com}{shinji.hirano@gmail.com}}
  \,and Vinayak Raj$^{a}$\footnote{
	e-mail:
	\href{mailto:vinayak.hep.th@gmail.com}{vinayak.hep.th@gmail.com}}}
\date{}

\maketitle

\thispagestyle{fancy}
\rhead{YITP-25-112}
\cfoot{}
\renewcommand{\headrulewidth}{0.0pt}

\vspace*{-1cm}
\begin{center}
%  \spa{0.5} \\
$^{a}${{\it School of Science, Huzhou University}}
\\ {{\it Huzhou 313000, Zhejiang, China}}
  \spa{0.5} \\
$^b${{\it Center for Gravitational Physics and Quantum Information (CGPQI)}}
\\ {{\it  Yukawa Institute for Theoretical Physics, Kyoto University}}
\\ {{\it Kitashirakawa-Oiwakecho, Sakyo-ku, Kyoto 606-8502, Japan}}
\spa{0.5}  

\end{center}

\begin{abstract}
We study correlators in two-dimensional $T\bar{T}$-deformed conformal field theories by interpreting the $T\bar{T}$ deformation as a coupling to two-dimensional gravity. To demonstrate the utility of the massive gravity framework as a particular realization of the gravitational interpretation, we show how the $T\bar{T}$-deformed correlators at finite coupling can be computed by adopting a judicious parametrization of the 2D metric and a preferred choice of zweibeins. To illustrate how this method works in practice, we compute the leading logarithmic contributions to two- and three-point functions to all orders in the $T\bar{T}$ coupling, reproducing a known result while producing new findings. This framework generalizes the random geometry approach to finite coupling.
\end{abstract}

\renewcommand{\thefootnote}{\arabic{footnote}}
\setcounter{footnote}{0}

\newpage

\tableofcontents

%\newpage
%%%%%%%%%%%%%%%%%%%%%%%%%%%%%%%%%%%%%%%%%%%%%%%%%%%%%%%%%%

\section{Introduction}
\label{sec:Introduction}

The $T\bar{T}$ deformation is a rare example of an irrelevant operator that leads to a well-defined and solvable quantum field theory~\cite{Zamolodchikov:2004ce,Smirnov:2016lqw,Cavaglia:2016oda}. Typically, irrelevant deformations introduce uncontrolled ultraviolet (UV) behavior, as their couplings grow with energy and signal the breakdown of the effective field theory. In most cases, this necessitates a UV completion. Remarkably, $T\bar{T}$-deformed theories avoid this fate: they remain under nonperturbative control and, moreover, introduce a built-in length scale that alters the short-distance structure of the theory. This suggests the possibility of qualitatively new short-distance physics -- distinct from that of conventional local quantum field theories (QFTs) -- and opens a window onto novel forms of nonlocality that may reflect features of quantum gravity.

A useful perspective interprets the deformation as coupling the theory to two-dimensional gravity~\cite{Dubovsky:2017cnj, Dubovsky:2018bmo, Tolley:2019nmm}, framing the $T\bar{T}$-deformed theory itself as a two-dimensional gravitational theory with dynamical geometry. In this paper, we build on this framework to study correlation functions in $T\bar{T}$-deformed conformal field theories (CFTs) at finite coupling, using the massive gravity formulation developed by Tolley~\cite{Tolley:2019nmm}, which extends the approach of~\cite{Hirano:2020nwq} from infinitesimal to finite deformation.\footnote{See also~\cite{Tsolakidis:2024wut} for a related discussion.} One of the key ingredients in our analysis is a judicious parametrization of the two-dimensional metric and a preferred choice of zweibeins, which together enable a systematic and geometrically transparent treatment of correlation functions in the deformed theory.

Within this setup, we compute leading logarithmic corrections to two-point functions order by order in the $T\bar{T}$ coupling and derive explicit all-order expressions for general scaling dimensions. Our method efficiently reproduces known perturbative results~\cite{Cardy:2019qao, Hirano:2020nwq, Hirano:2024eab} and extends them in a unified and geometrically motivated framework. We then generalize the computation to three-point functions and obtain a new closed-form expression that captures the leading logarithmic corrections to all orders in perturbation theory. Finally, we offer a brief preview of how these results connect to a nonperturbative completion of the two-point correlators, to be developed in a forthcoming companion paper~\cite{HR_NP_Planck}.

The remainder of this paper is organized as follows. In Section~\ref{sec:2Dgravity}, we review the massive gravity formulation of the $T\bar{T}$ deformation and discuss our choice of parametrization for the metric and zweibein, which plays a central role in setting up the computation of correlators. Section~\ref{sec:2pt} is devoted to the computation of leading logarithmic corrections to two-point functions, carried out to all orders in perturbation theory. We also provide a brief preview of the nonperturbative completion of two-point correlators. In Section~\ref{sec:3pt}, we extend the analysis to three-point functions and derive a closed-form expression for the all-order leading-logarithmic contributions. We conclude in Section~\ref{sec:discussion} with a summary of our findings and future directions.

{\bf Note added}: While this work engages with a technical theme closely related to that of~\cite{Aharony:2023dod}, it differs in important ways in both methodology and findings. Our analysis builds on the massive gravity formulation of~\cite{Tolley:2019nmm}, whereas their approach is based on a closely related but slightly different description developed in~\cite{Dubovsky:2017cnj, Dubovsky:2018bmo}, namely, fixed background versus dynamical coordinates. Our perspective also differs from theirs in how we interpret the role and significance of the gravitational description. We regard the $T\bar{T}$-deformed CFT as a conformal field theory defined on a fluctuating geometry and aim to probe the effects of dynamical geometry. To this end, we study standard local operators on the ``worldsheet,'' whereas their focus is on local operators in the ``target space,'' which appear non-local from the worldsheet point of view. Their main results on the short-distance behavior of correlators differ from ours -- though they bear some qualitative similarities -- which are briefly presented at the end of Section~\ref{sec:all_order} and will be explored in more detail in a forthcoming companion paper~\cite{HR_NP_Planck}.\footnote{The short-distance results reported in~\cite{Cui:2023jrb,Chen:2025jzb} are consistent with those of~\cite{Aharony:2023dod}, whose method was further applied to correlators on a torus in~\cite{Barel:2024dgv}.}

%%%%%%%%%%%%%%%%%%%%%%%%%%%%%%%%%%%%%%%%%%%%%%%%%%%%%%%
\section{$T\Tb$-deformed correlators from a 2D gravity description}
\label{sec:2Dgravity}

As outlined in the Introduction, we investigate $T\bar{T}$-deformed correlators within the massive gravity formulation proposed by Tolley~\cite{Tolley:2019nmm}, with the goal of advancing the viewpoint that the $T\bar{T}$ deformation defines a quantum gravitational theory endowed with a built-in length scale $\ell_P=\sqrt{|\mu|}$, where $\mu$ is the $T\Tb$ coupling. To make this perspective concrete and demonstrate its utility, we work within a two-dimensional gravity description of the $T\Tb$-deformed CFT, characterized by the action\footnote{The convention for the Levi-Civita symbols is such that $\epsilon^{01}=\epsilon_{01}=1$.}
\begin{align}\label{FT_action}
S_{T\Tb}[e, f]
=S_{\rm CFT}[e]+{1\over 2\mu}\int d^2x\epsilon^{ij}\epsilon_{ab}\left(e^a_{i}-f^a_{i}\right)\left(e^b_{j}-f^b_{j}\right)
\equiv S_{\rm CFT}[e]+S_{\rm mG}[e, f]\ .
\end{align}
Here, $e$ and $f$ are zweibeins corresponding to the two-dimensional metrics
$ds_{\rm CFT}^2 = h_{ij} dx^i dx^j$ and
$ds_{T\bar{T}}^2 = \gamma_{ij} dx^i dx^j$, respectively. 
Importantly, the metric $h_{ij}$ (or equivalently, the zweibein $e^a_i$) is dynamical, rendering the theory a form of two-dimensional quantum gravity.
The presence of the two metrics reflects two complementary descriptions of the
$T\bar{T}$-deformed theory: either as a deformed CFT living on a fixed
(undeformed) background with metric $\gamma_{ij}$, or as an undeformed CFT
living on a dynamically fluctuating geometry with metric $h_{ij}$. This perspective was proposed in~\cite{Conti:2018tca, Cardy:2019qao} and further developed, for example, in~\cite{Hirano:2024eab}.

The observables of the $T\bar{T}$-deformed CFT are related to those of the undeformed CFT through the generating functional of the deformed theory, given by
\begin{align}
Z_{T\Tb}[f, J]=\int{\cal D}e^a_i e^{-S_{\rm mG}[e, f]}Z_{\rm CFT}[e, J]\ ,
\end{align}
where $J$ collectively denotes the sources for the operators in the theory. This functional solves the diffusion equation or ``Schr\"odinger equation,'' 
\begin{align}\label{qFlow}
{\del\over \del\mu}Z_{T\Tb}[f,J]=\int d^2x \lim_{y\to x}{1\over 2}\epsilon_{ij}\epsilon^{ab}{\delta^2Z_{T\Tb}[f, J]\over \delta f^a_{i}(y)\delta f^b_{j}(x)}
\end{align}
which arises from the random geometry interpretation of the $T\bar{T}$ deformation at infinitesimal coupling~\cite{Cardy:2018sdv}.

%%%%%%%%%%%%%%%%%%%%%%%%%%%%%%%%%%%%%%%%%%%%%%%%%%%%%%%
\subsection{Decomposition of the metric and zweibein}
\label{sec:decomposition}

To compute correlators within this massive gravity formulation, we follow the approach developed in~\cite{Hirano:2020nwq}. Exploiting the fact that any two-dimensional space is conformally equivalent to flat space, we parametrize the boundary metric $h_{ij}$ as
\begin{align} \label{metric_parametrization}
ds_{\rm CFT}^2 = h_{ij}\, dx^{i} dx^{j} = e^{2\Phi(x)} \delta_{ij}\,d(x^{i} + \alpha^{i}(x)) \,d(x^{j} + \alpha^{j}(x))\ .
\end{align}
The three unconstrained degrees of freedom of the two-dimensional metric are thus encoded in the Weyl and diffeomorphism components: $(\Phi, \alpha_1, \alpha_2)$.
The zweibein $e^a_{i}$ is not uniquely determined. However, as we will see, there appears to be a preferred choice,\footnote{One might naively consider the choice
\begin{align}\label{e_phi_alpha_naive}
e^a_{i}=\delta^a_{i}+\delta^a_{i}\Phi+\del_{i}\alpha^a+{\cal O}(\alpha^2, \alpha\Phi, \Phi^2)\ .
\end{align}
However, as discussed below, this choice fails to reproduce results consistent with the random geometry approach of~\cite{Hirano:2020nwq}.}
\begin{align}\label{e_phi_alpha}
e^a_{i}=\delta^a_{i}+\delta^a_{i}\Phi+{1\over 2}\delta^a_{k}(\del_{i}\alpha^{k}+\del^{k}\alpha_{i})
+{\cal O}(\alpha^2, \alpha\Phi, \Phi^2)\ .
\end{align}
A systematic expansion of the zweibein to higher orders follows from the requirement that the metric satisfies
\begin{align}
h_{ij}=\delta_{ab}e^a_{i}e^b_{j}=e^{2\Phi}\left(\delta_{ij}+\del_{i}\alpha_{j}+\del_{j}\alpha_{i}+\del_{i}\alpha^{k}\del_{j}\alpha_{k}\right)\ .
\end{align}
This yields
\begin{equation}
\begin{aligned}
e^a_{i}&=e^{\Phi}\biggl[\delta^a_{i}+ {1\over 2}\delta^a_{k}(\del_{i}\alpha^{k}+\del^{k}\alpha_{i})
-{1\over 8}\delta^a_{k}(\del_{i}\alpha^{l}+\del^{l}\alpha_i)(\del^{k}\alpha_{l}+\del_{l}\alpha^{k})
+{1\over 2}\delta^a_{k}\del_{i}\alpha^{l}\del^{k}\alpha_{l}\\
&\hspace{.4cm}+{1\over 2}\delta^a_{k}(\del_{i}\alpha^{l}+\del^{l}\alpha_{i})
\biggl({1\over 8}(\del^{k}\alpha^{m}+\del^{m}\alpha^{k})(\del_{l}\alpha_{m}+\del_{m}\alpha_{l})
-{1\over 2}\del^{k}\alpha^{m}\del_{l}\alpha_{m}\biggr)
+{\cal O}(\alpha^4)\biggr].
\end{aligned}
\end{equation}
To isolate the intrinsic Weyl mode $\phi$ from the contribution induced by conformal transformations, we decompose the total Weyl factor $\Phi$ as
\begin{align}
2\Phi = 2\phi - \ln\det(\delta^{i}_{j}+\del_{j}\alpha^{i}) 
=2\phi -\ln\left(1+\del_{\lambda}\alpha^{\lambda}
+{1\over 2}\left((\del_{k}\alpha^{k})^2-\del_{k}\alpha^{l}\del_{l}\alpha^{k}\right)\right)\ .
\end{align}
With this parametrization in place, for the $T\bar{T}$-deformed CFT on $\mathbb{R}^2$ -- that is, with $\gamma_{ij} = \delta_{ij}$ and $f^a_i = \delta^a_i$ -- the massive gravity action becomes
\begin{equation}
\begin{aligned}\label{mGaction}
S_{\rm mG}[e,f]={1\over\mu}\int_{\p AdS_3} d^2x\left(\phi^2+{1\over 4}\alpha^{i}\Box\alpha_{i}+{\cal O}(\alpha^3,\alpha^2\phi, \alpha\phi^2, \phi^3)\right)
\equiv S_{\rm mG}(\phi, \alpha)\ ,
\end{aligned}
\end{equation}
which reproduces the Hubbard-Stratonovich action found in~\cite{Hirano:2020nwq}. We note that had we instead used the naive zweibein ansatz~\eqref{e_phi_alpha_naive}, the resulting action would have been
\begin{equation}
\begin{aligned}\label{mGwrong}
S_{\rm mG}[e,f]={1\over\mu}\int_{\p AdS_3} d^2x\left(\phi^2+{1\over 4}\alpha^{i}\del_{i}\del_{j}\alpha^{j}+{\cal O}(\alpha^3,\alpha^2\phi, \alpha\phi^2, \phi^3)\right)\ ,
\end{aligned}
\end{equation}
which fails to match the known result.

%%%%%%%%%%%%%%%%%%%%%%%%%%%%%%%%%%%%%%%%%%%%%%%%%%%%%%%
\subsection{Prescription for $T\Tb$-deformed correlators}
\label{sec:correlators}

The decomposition of the metric in \eqref{metric_parametrization} provides an intuitive picture of how the $T\bar{T}$ deformation modifies correlators:
the insertion points of operators fluctuate as a result of a dynamical coordinate transformation, weighted by the massive gravity action \eqref{mGaction}.\footnote{Here, ``dynamical'' means that $\alpha^i(x)$ is integrated over, i.e., treated as a fluctuating field.}

As we will see, all nontrivial spatial dependence arises from the two-dimensional propagator $\Box^{-1}={1\over 2\pi}\ln(|x-x'|/\varepsilon)$ through the Gaussian action $\alpha^i\Box\alpha_i$. This implies that the conformal transformations satisfying $\Box\alpha_i=0$ do not contribute nontrivially to the $T\Tb$-deformed correlators.
Likewise, the intrinsic Weyl mode $\phi$ does not generate nontrivial contributions either.\footnote{In contrast, when stress tensors are present, the Weyl mode $\phi$ contributes nontrivially through the conformal anomaly action and plays a significant role~\cite{Hirano:2020ppu}.}
In light of this and in the absence of nontrivial contributions from conformal transformations and Weyl scaling, the nontrivial part of the $T\Tb$-deformed correlators may be expressed as
\begin{align}\label{npt_FT}
\left\langle \prod_{A=1}^n{\cal O}_{\Delta_A}(x_A)\right\rangle_{T\Tb}={\cal N}^{-1}\prod_{i=1}^2{\cal D}\alpha^{i}
\left\langle\prod_{B=1}^n {\cal O}_{\Delta_B}(x_B+\alpha(x_B))\right\rangle_{\rm CFT}\!\! e^{-S_{\rm mG}(\phi=0, \alpha)}
\end{align}
where ${\cal N}$ is a normalization constant chosen such that the correlators reduce to their CFT counterparts in the $\mu \to 0$ limit. Here, the conformal dimension $\Delta_A$ refers to the sum of the holomorphic and anti-holomorphic scaling dimensions.
Importantly, this expression omits both the $\phi$-integration and the associated dependence, as well as the would-be Jacobian factors $\prod_{A=1}^nJ(x_A)$, with $J(x_A)=\left(e^{2\Phi(x_A)}\det(\delta^{i}_{j}+\del_{j}\alpha^{i}(x_A))\right)^{\Delta_A/2}$, which arise from Weyl scaling and conformal transformations.
While it is not obvious how to determine the appropriate path-integral measure, we simply adopt the most straightforward choice. As we will see below, potential Jacobian corrections to the measure do not affect, at least, the leading logarithmic contribution to the correlators.

Our primary aim is to demonstrate the utility of the massive gravity framework as a concrete realization of the quantum gravity interpretation of the $T\bar{T}$ deformation. To this end, we begin with a more modest objective: reproducing a known result -- the leading logarithmic corrections to two-point correlators obtained in~\cite{Cardy:2019qao} (see also \cite{Hirano:2020nwq, Hirano:2024eab}):
\begin{equation}
\begin{aligned}\label{2pt_all_orders}
\langle {\cal O}_{\Delta}(x_1){\cal O}_{\Delta}(x_2)\rangle^{\text{leading-log}}_{\mu}
=\sum_{n=0}^{\infty}(-1)^n{(4\mu)^n\over n!\pi^{n}}\prod_{k=0}^{n-1}(\Delta+k)^2
{\ln^n(|x_{12}|/\varepsilon)\over |x_{12}|^{2(\Delta+n)}}\ .
\end{aligned}
\end{equation}
As will be elaborated below, the situation is simpler than it may initially appear. Specifically, the higher-order corrections in the massive gravity action~\eqref{mGaction} always yield subleading contributions in powers of the logarithm, as they involve derivatives acting on the leading logarithmic terms. Therefore, to compute the leading logarithmic corrections, it suffices to consider the Gaussian action alone. For the same reason, as noted above, potential Jacobian corrections to the integration measure also contribute only at subleading order.

%%%%%%%%%%%%%%%%%%%%%%%%%%%%%%%%%%%%%%%%%%%%%%%%%%%%%%%
\section{Two-point correlators}
\label{sec:2pt}

As we discuss later, the prescription~\eqref{npt_FT} allows the two-point correlators to be computed to all orders in perturbation theory, reproducing the known result~\eqref{2pt_all_orders}.\footnote{In fact, the expression admits a nonperturbative completion, as discussed in a companion paper~\cite{HR_NP_Planck}.}
Nevertheless, it is instructive to illustrate how the $T\bar{T}$ corrections can be computed systematically, order by order in perturbation theory, particularly with future applications in mind.\footnote{We note several earlier related works:~\cite{He:2019vzf,He:2020qcs} present a first-order perturbative analysis of two- and four-point functions, with applications to entanglement and quantum chaos;~\cite{He:2023kgq} introduces a systematic recursive framework for computing correlators to arbitrary order. See also~\cite{He:2025ppz} for a review of these developments.}

%%%%%%%%%%%%%%%%%%%%%%%%%%%%%%%%%%%%%%%%%%%%%%%%%%%%%%%
\subsection{Perturbative method}
\label{sec:perturbative}

Our strategy is to evaluate the path integral using the Gaussian weight -- i.e., the quadratic part of the massive gravity action~\eqref{mGaction} -- by treating all remaining $\alpha_i$-dependent terms as perturbative insertions. Note that the $n$-th order correction corresponds to the ${\cal O}(\alpha^{2n})$ contribution, as seen by rescaling $\alpha^{i} \to \sqrt{\mu}\alpha^{i}$. Due to the Gaussian weight, only even powers of $\alpha$ contribute, and hence only even-$n$ corrections survive. 

%%%%%%%%%%%%%%%%%%%%%%%%%%%%%%%%%%%%%%%%%%%%%%%%%%%%%%%%%%%
\subsubsection{First-order consistency check (warm-up)}\label{sec:1storder}

The terms in the integrand of \eqref{npt_FT} relevant to the first-order correction are therefore given by
\begin{equation}
\begin{aligned}\label{quadratic}
\hspace{-.0cm}
{1\over \left|x_{12}+\alpha_{12}\right|^{2\Delta}}&={1\over |x_{12}|^{2\Delta}}+{\cal O}(\alpha)
-{\Delta\alpha_{12}^{i}\alpha_{12}^{j}\over |x_{12}|^{2\Delta+2}}\left(\delta_{ij}-{(2\Delta+2)(x_{12})_{i}(x_{12})_{j} \over |x_{12}|^2}\right)\ ,
\end{aligned}
\end{equation}
where we have defined $x^i_{12}=x^i_1-x^i_2$, $\alpha^i_A=\alpha^i(x_A)$, and $\alpha^i_{12}=\alpha^i_1-\alpha^i_2$.
This expression is to be integrated with the Gaussian weight. A standard technique for performing such Gaussian integrals is to introduce a source term:
\begin{equation}
\begin{aligned}\label{standard_technique}
\hspace{-.3cm}
I[J]&=\int\prod_{i}{\cal D}\alpha^{i}F[\alpha^{j}(y)]e^{-\int d^2x{1\over 4\mu}\alpha_i(x)\Box\alpha^i(x)+J_i(x)\alpha^i(x)}\\
&=F\left[{\delta\over \delta J^{k}(y)}\right]\int\prod_{i}{\cal D}\alpha^{i}
e^{-\int d^2x\left[{1\over 4\mu}(\alpha_{i}-2\mu\int d^2x'\Box^{-1}J_{i})
\Box(\alpha^{i}-2\mu\int d^2x'\Box^{-1}J^{i})+\mu\int d^2x' J_{i}\Box^{-1}J^{i}\right]}\\
&={\cal N}F\left[{\delta\over \delta J^{k}(y)}\right]e^{\mu\int d^2x\int d^2x' J(x)_{i}\Box^{-1}(x,x')J(x')^{i}}
\end{aligned}
\end{equation}
where $\Box^{-1}(x, x')={1\over 2\pi}\ln(|x-x'|/\varepsilon)$ is the Green’s function in two dimensions, with $\varepsilon$ an arbitrary length scale, and ${\cal N}$ is the regularized value of the infinite-dimensional Gaussian integral. The source $J_i(x)$ is turned off at the end of the calculation.

Let us now comment on a convergence subtlety. The Gaussian integral in~\eqref{standard_technique} is convergent only for positive $\mu>0$.
However, when considering perturbative contributions, alternative techniques -- such as conformal perturbation theory or the use of dynamical coordinate transformations~\cite{Hirano:2024eab} -- are available. The result~\eqref{2pt_all_orders} supports the validity of a naive analytic continuation from positive to negative $\mu$. 
Since our focus here is on perturbation theory, we adopt this analytic continuation throughout.%
\footnote{This assumption does not hold in a nonperturbative context; see the companion paper~\cite{HR_NP_Planck} for details.}

It is now straightforward to compute the first-order correction using our prescription~\eqref{npt_FT}. Equation~\eqref{quadratic} provides the relevant integrand for the Gaussian path integral. When two $\alpha$ fields are evaluated at coincident points, the resulting terms are divergent, involving structures such as $\delta^2(0)$ and its derivatives, e.g., $\partial^m \delta^2(0)$. These divergences can be renormalized, for instance, via point-splitting regularization. We therefore focus on terms in which one $\alpha$ is evaluated at $x_1$ and the other at $x_2$:
\begin{equation}
\begin{aligned}
\langle {\cal O}_{\Delta}(x_1){\cal O}_{\Delta}(x_2)\rangle^{\text{leading-log}}_{T\Tb}&={1\over |x_{12}|^{2\Delta}}
+\hat{X}_{ij}{\delta^2 e^{\mu\int d^2x\int d^2x' J_i(x)\Box^{-1}(x,x')J^i(x')}\over\delta J_{i}(x_1)\delta J_{j}(x_2)}\Biggr|_{J=0}\\
&={1\over |x_{12}|^{2\Delta}}-{4\mu\Delta^2\ln(|x_{12}|/\varepsilon)\over \pi |x_{12}|^{2\Delta+2}}\ ,
\end{aligned}
\end{equation}
where we have defined the tensor
\begin{equation}
\begin{aligned}\label{X_ij}
\hat{X}_{ij}&={2\Delta\over |x_{12}|^{2\Delta+2}}\left(\delta_{ij}-{(2\Delta+2)(x_{12})_{i}(x_{12})_{j} \over |x_{12}|^2}\right)\ .
\end{aligned}
\end{equation}
This result correctly reproduces the known expression obtained in~\cite{Kraus:2018xrn, Cardy:2019qao}. We have thus confirmed that our prescription is consistent with the random geometry approach proposed in~\cite{Hirano:2020nwq}, as expected.

%%%%%%%%%%%%%%%%%%%%%%%%%%%%%%%%%%%%%%%%%%%%%%%%%%%%%%%%%%%
\subsubsection{Second-order consistency check (nontrivial test)}\label{sec:2ndorder}

As remarked earlier, the prescription~\eqref{npt_FT} provides the finite-coupling extension of the approach developed in~\cite{Hirano:2020nwq}. The first nontrivial check is therefore to verify whether the second-order correction to two-point functions is correctly reproduced. As noted at the beginning of Section~\ref{sec:1storder}, the $\mathcal{O}(\mu^2)$ contribution corresponds to quartic order in $\alpha$. Since the leading logarithmic behavior arises from terms without derivatives acting on $\alpha$, the dominant contribution comes from the $\mathcal{O}(\alpha^4)$ terms in the Taylor expansion of the undeformed CFT two-point function (see Appendix~\ref{app:details} for details). In other words, as previously noted, the non-Gaussian higher-order terms in the massive gravity action~\eqref{mGaction} can be neglected entirely when computing the leading logarithmic corrections.

Focusing on the leading logarithmic contribution, we obtain
\begin{equation}
\begin{aligned}\label{2pt_second}
\langle {\cal O}_{\Delta}(x_1){\cal O}_{\Delta}(x_2)\rangle^{\text{leading-log}}_{T\Tb}&={1\over |x_{12}|^{2\Delta}}
+\hat{X}_{ijkl}{\delta^4e^{\mu\int d^2x\int d^2x' J_i(x)\Box^{-1}(x,x')J^i(x')}\over\delta J_{i}(x_{12})\delta J_{j}(x_{12})\delta J_{k}(x_{12})\delta J_{l}(x_{12})}\Biggr|_{J=0}\ ,
\end{aligned}
\end{equation}
where we have defined $\delta/\delta J_{i}(x_{12})\equiv \delta/\delta J_{i}(x_{1})-\delta/\delta J_{i}(x_{2})$. The tensor $\hat{X}_{ijkl}$ is given by
\begin{equation}
\begin{aligned}\label{X_ijkl}
\hat{X}_{ijkl}&={\Delta(\Delta+1)\over 3! |x_{12}|^{2\Delta+4}}
\biggl[\delta_{ij}\delta_{kl}+\delta_{ik}\delta_{jl}+\delta_{il}\delta_{jk}\\
&-{2(\Delta+2) \over |x_{12}|^2}\biggl(\delta_{il}(x_{12})_{j}(x_{12})_{k}+\delta_{jl}(x_{12})_{i}(x_{12})_{k}+\delta_{kl}(x_{12})_{i}(x_{12})_{j} \\
&+\delta_{ij}(x_{12})_{k}(x_{12})_{l}+\delta_{ik}(x_{12})_{j}(x_{12})_{l}+\delta_{jk}(x_{12})_{i}(x_{12})_{l}  \biggr)\\
&+{4(\Delta+2)(\Delta+3)(x_{12})_{i}(x_{12})_{j}(x_{12})_{k}(x_{12})_{l}  \over |x_{12}|^4}\biggr]\ .
\end{aligned}
\end{equation}
If one includes the non-Gaussian higher-order terms from the massive gravity action~\eqref{mGaction}, they would modify $\hat{X}_{ijkl}$ by introducing additional differential operators acting on $\Box^{-1}$. However, such terms only affect subleading contributions and do not alter the leading logarithmic behavior of the correlator.

To evaluate the second term in~\eqref{2pt_second}, we must work out the tensor contractions and combinatorics. Carrying out this analysis (see Appendix~\ref{app:details} for details), we find
\begin{align}
\label{TTbar_2nd_2pt}
\langle {\cal O}_{\Delta}(x_1){\cal O}_{\Delta}(x_2)\rangle^{\text{leading-log}}_{T\Tb}
&={1\over |x_{12}|^{2\Delta}}+{8\mu^2\Delta^2(\Delta+1)^2\over \pi^2}{\ln^2(|x_{12}|/\varepsilon)\over |x_{12}|^{2\Delta+4}}\ .
\end{align}
This expression precisely matches the second-order term in~\eqref{2pt_all_orders}, thereby confirming the validity of our prescription.

%%%%%%%%%%%%%%%%%%%%%%%%%%%%%%%%%%%%%%%%%%%%%%%%%%%%%%%
\subsection{All-order perturbative analysis}
\label{sec:all_order}

As we now explain, our prescription~\eqref{npt_FT} enables a much simpler computation of two-point correlators, bypassing the need for brute-force perturbative methods.
A key technical step is to Fourier transform the undeformed CFT two-point function using the identity
\begin{align}\label{Fourier_Transform}
\int_{-\infty}^{\infty}d^dk e^{i\vec{k}\cdot\vec{x}} |k|^a
=\int_{-\infty}^{\infty}d^dk e^{i\vec{k}\cdot\vec{x}}{1\over \Gamma\left(-{a\over 2}\right)}\int_0^{\infty}dt t^{-{a\over 2}-1} e^{-|k|^2t} 
&={\pi^{d\over 2}2^{a+d}\Gamma\left({a+d\over 2}\right)\over \Gamma\left(-{a\over 2}\right)}|x|^{-(a+d)}\ ,
\end{align}
where in the second line we performed the Gaussian integral over $\vec{k}$ and evaluated the $t$-integral using the definition of the Gamma function after a change of variables.
In our application, we have $d=2$, $a=2(\Delta-1)$, and $\vec{x}=\vec{x}_{12}+\vec{\alpha}_{12}$. Hence, the computation reduces to performing Gaussian integrals -- for generic values of $\Delta$, excluding $\Delta\in \mathbb{Z}_+$:
\begin{equation}
\begin{aligned}\label{2pt_FT_leading_log_FT}
\langle {\cal O}_{\Delta}(x_1){\cal O}_{\Delta}(x_2)\rangle^{\rm leading{\text -}log}_{T\Tb}
&={\cal N}^{-1}{ \Gamma\left(-\Delta+1\right)\over \pi 2^{2\Delta}\Gamma\left(\Delta\right)}\int \prod_{i}{\cal D}\alpha^{i}\int_{-\infty}^{+\infty}d^2k
e^{ik_{i}(x_{12}+\alpha_{12})^{i}}\\
&\times |k|^{2(\Delta-1)}e^{-{1\over 4\mu}\int d^2x\alpha_{i}(x)\Box\alpha^{i}(x)}\\
&={\cal N}^{-1}{ \Gamma\left(-\Delta+1\right)\over \pi 2^{2\Delta}\Gamma\left(\Delta\right)}\int \prod_{i}{\cal D}\alpha^{i}\int_{-\infty}^{+\infty}d^2k
e^{i\vec{k}\cdot\vec{x}_{12}}|k|^{2(\Delta-1)}\\
&\times e^{-\int d^2x{1\over 4\mu}(\alpha_{i}(x)-2\mu K_{i}(x))\Box(\alpha^{i}(x)-2\mu K^{i}(x))+{\mu\over\pi}|k|^2\ln(|x_{12}|/\varepsilon)+{\rm div.}}\\
&={ \Gamma\left(-\Delta+1\right)\over \pi 2^{2\Delta}\Gamma\left(\Delta\right)}\int_{-\infty}^{+\infty}d^2k|k|^{2(\Delta-1)}e^{i\vec{k}\cdot\vec{x}_{12}+{\mu\over\pi}|k|^2\ln(|x_{12}|/\varepsilon)+{\rm div.}}\ .
\end{aligned}
\end{equation}
Here, we have defined
\begin{align}
K_{i}(x)\equiv ik_{i}\int d^2x'\left(\delta^2(x'-x_1)-\delta^2(x'-x_2)\right)\Box^{-1}(x',x)
\end{align}
and ``div.'' denotes a contact divergence arising from coincident points, similar to those encountered in the perturbative calculation. This divergence is renormalized, for example, via point-splitting regularization.

Before proceeding with the evaluation of the integrals in~\eqref{2pt_FT_leading_log_FT}, let us note two subtleties related to convergence. The first concerns the Gaussian integrals over $\alpha_i$ in the second line of the expression, as discussed below~\eqref{standard_technique}. As argued there, we can adopt a naive analytic continuation in $\mu$ as long as we restrict ourselves to perturbative calculations. The second issue involves the convergence of the $k_i$-integrals appearing in the final expression. 
At large distances $|x_{12}| > \varepsilon$, above the reference scale $\varepsilon$ of our choice, the integrals converge only for positive $\mu > 0$. In contrast, at short distances $|x_{12}| < \varepsilon$, they converge for negative $\mu < 0$. However, since our focus is on the perturbative expansion, we treat the final expression as a formal (non-convergent) series and set aside these convergence subtleties. We then find
\begin{equation}
\begin{aligned}\label{2pt_all_orders_mG}
\hspace{-.4cm}
\langle {\cal O}_{\Delta}(x_1){\cal O}_{\Delta}(x_2)\rangle^{\text{leading-log}}_{T\Tb}&
=\sum_{n=0}^{\infty}{\mu^n\Gamma\left(-\Delta+1\right)\over n!\pi^{n+1} 2^{2\Delta}\Gamma\left(\Delta\right)}
\int_{-\infty}^{+\infty}d^2k|k|^{2(\Delta-1+n)}e^{i\vec{k}\cdot\vec{x}_{12}}\ln^n(|x_{12}|/\varepsilon)\\
&=\sum_{n=0}^{\infty}(-1)^n{(4\mu)^n\over n!\pi^{n}}\prod_{k=0}^{n-1}(\Delta+k)^2
{\ln^n(|x_{12}|/\varepsilon)\over |x_{12}|^{2(\Delta+n)}}\ .
\end{aligned}
\end{equation}
This expression precisely matches the known result~\eqref{2pt_all_orders}.

We now comment on the special case \(\Delta \in \mathbb{Z}_+\). While this case requires additional care, the final result coincides with that obtained for generic (non-integer) values of $\Delta$.
In this case, the exponent \(a = 2(\Delta - 1)\) in the Fourier transform~\eqref{Fourier_Transform} becomes a non-negative even integer, and the factor \(\Gamma(-a/2)\) in the denominator diverges. As a result, the standard position-space expression vanishes. To resolve this, we differentiate the Fourier integral with respect to \(a\), which yields
\begin{equation}
\begin{aligned}
\int_{-\infty}^{\infty}d^dk\, e^{i\vec{k}\cdot\vec{x}}\, |k|^a\ln|k|
&= \pi^{d/2} 2^{a+d-1} \Gamma\left({a+d \over 2}\right) {\Gamma'\left(-{a \over 2}\right) \over \Gamma\left(-{a \over 2}\right)^2} |x|^{-(a+d)} \\
&= (-1)^{{a \over 2}+1} \pi^{d/2} 2^{a+d-1} \Gamma\left({a+d \over 2}\right)^2 |x|^{-(a+d)} ,
\end{aligned}
\end{equation}
where we used the asymptotic behavior $\Gamma(-a/2) \sim {(-1)^n / [n!(n - a/2)]}$ and $\Gamma'(-a/2) \sim {(-1)^{n+1} / [n!(n - a/2)^2]}$ near the pole $a/2 = n\in\mathbb{Z}_+\cup\{0\}$. Note that this is the only surviving contribution among various terms, as $1/\Gamma(-a/2) \propto (n - a/2)$ vanishes in the limit.
Nevertheless, applying this version of the Fourier transform in~\eqref{2pt_FT_leading_log_FT} yields the same result as in the generic (non-integer) case.

Finally, we provide a preview of the nonperturbative completion of the two-point correlators, to be detailed in an upcoming companion paper~\cite{HR_NP_Planck}. Identifying the reference scale $\varepsilon$ with the ``Planck scale'' $\ell_P \equiv \sqrt{|\mu|}$, the nonperturbative correlators take the form
\begin{align}\label{NP2pt_corrected}
\hspace{-.4cm}
\langle {\cal O}_{\Delta}(x_1){\cal O}_{\Delta}(x_2)\rangle^{\text{leading-log}}_{T\Tb}
&=
\begin{cases}
\dfrac{(-Z)^{-\Delta}}{|x_{12}|^{2\Delta}}  U\left(\Delta, 1, -\dfrac{1}{Z}\right) & \text{for } Z < 0 \\[1ex]
\dfrac{\Gamma(1-\Delta) Z^{-\Delta}}{|x_{12}|^{2\Delta}} e^{-{1\over Z}}  \mbox{}_1F_1\left(1-\Delta; 1; \dfrac{1}{Z}\right) & \text{for } Z > 0
\end{cases}
\end{align}
for generic (non-integer) values of $\Delta$, where $U(a,b,z)$ is the Tricomi confluent hypergeometric function of the second kind, and
\[
Z \equiv -\frac{4\mu}{\pi |x_{12}|^2} \ln\left(\frac{|x_{12}|}{\ell_P}\right).
\]
A similar, though slightly modified, expression applies to the special case $\Delta \in \mathbb{Z}_+$.
Both functions in~\eqref{NP2pt_corrected} admit the asymptotic expansion~\eqref{2pt_all_orders_mG} at large distances $|x_{12}| \gg \ell_P$, though the latter takes the form of a trans-series that includes an instanton-like sector.
At short distances $|x_{12}| \ll \ell_P$, correlations are typically suppressed. While the short-distance behavior shares qualitative features with the findings of~\cite{Aharony:2023dod,Cui:2023jrb,Chen:2025jzb}, it differs in quantitative detail.
 
%%%%%%%%%%%%%%%%%%%%%%%%%%%%%%%%%%%%%%%%%%%%%%%%%%%%%%%
\section{Three-point correlators}
\label{sec:3pt}

We now extend the all-orders computation of leading logarithmic corrections to three-point correlators using our prescription~\eqref{npt_FT}. As in the two-point case, it suffices to work with the stripped-down version of the massive gravity description:
\begin{align} \label{3pt_FT_leading_log}
\hspace{-.35cm}
\left\langle \prod_{A=1}^3 \mathcal{O}_{\Delta_A}(x_A) \right\rangle^{\text{leading-log}}_{T\bar{T}}
= {\cal N}^{-1} \int \prod_{i} \mathcal{D} \alpha^{i}
\left\langle \prod_{A=1}^3 \mathcal{O}_{\Delta_A}(x_A + \alpha(x_A)) \right\rangle_{\text{CFT}}
e^{- \frac{1}{4\mu} \int d^2x , \alpha_i \Box \alpha^i},
\end{align}
where the CFT three-point function takes the standard form
\begin{align}
\left\langle \prod_{A=1}^3 \mathcal{O}_{\Delta_A}(X_A) \right\rangle_{\text{CFT}}
= \frac{C_{\Delta_1\Delta_2\Delta_3}}{|X_{12}|^{\Delta_1 + \Delta_2 - \Delta_3} |X_{23}|^{\Delta_2 + \Delta_3 - \Delta_1} |X_{13}|^{\Delta_3 + \Delta_1 - \Delta_2}}\ .
\end{align}

Using the Fourier transform~\eqref{Fourier_Transform}, the integrals over $\alpha_i$ in~\eqref{3pt_FT_leading_log} become Gaussian. 
For convenience, we introduce the notation $\delta_C=(\Delta_A+\Delta_B-\Delta_C)/2$ where $A, B, C$ are all distinct. The CFT correlator for generic values of $\Delta$ then takes the form
\begin{equation}
\begin{aligned}
\hspace{-.0cm}
&\left\langle\prod_{A=1}^3 {\cal O}_{\Delta}(x_A+\alpha_A)\right\rangle_{\!\rm CFT}
= {\cal N}^{-1}C_{\Delta_1\Delta_2\Delta_3}\\
&\hspace{3.0cm}\times \!\prod_{(A,B,C)={\rm cycl}(1,2,3)}\!{\Gamma\left(-\delta_{A}+1\right)\over \pi 2^{2\delta_{A}}\Gamma\left(\delta_{A}\right)}\int_{-\infty}^{+\infty}d^2k_A
e^{i\vec{k}_A\cdot(\vec{x}_{BC}+\vec{\alpha}_{BC})}|k_A|^{2(\delta_{A}-1)}
\end{aligned}
\end{equation}
where ${\rm cycl}(1,2,3)$ denotes cyclic permutations of 1, 2, and 3.
Completing the square in $\alpha_i$, the exponent becomes
\begin{equation}
\begin{aligned}
&\int d^2x\biggl[-{1\over 4\mu}(\alpha_{i}-K_{i})\Box(\alpha^{i}-K^{i}) + {\rm div.}\\
&+\sum_{(A,B,C)={\rm cycl}(1,2,3)}\left(i\vec{k}_A\cdot\vec{x}_{BC}+{\mu\over 2\pi}2|k_A|^2\ln(|x_{BC}|/\varepsilon)
+\vec{k}_B\cdot \vec{k}_C\ln{\varepsilon|x_{BC}|\over |x_{AB}||x_{AC}|}\right)
\end{aligned}
\end{equation}
Here, ``div.'' denotes again a contact divergence, and we have defined
\begin{equation}
\begin{aligned}
K^{i}(x)&=2i \mu\!\!\!\!\sum_{(A,B,C)={\rm cycl}(1,2,3)}\!\!\!\!k_A^{i}\int d^2x'\left(\delta^2(x'-x_B)-\delta^2(x'-x_C)\right)\Box^{-1}(x',x)\ .
\end{aligned}
\end{equation}
After (wavefunction) renormalization via point-splitting, we find
\begin{equation}
\begin{aligned}
\left\langle \prod_{A=1}^3{\cal O}_{\Delta_A}(x_A)\right\rangle^{\text{leading-log}}_{T\Tb}
&\!\!\!\!\!\!\!=C_{\Delta_1\Delta_2\Delta_3}\!\!\!\!\!\prod_{(A,B,C)={\rm cycl}(1,2,3)}\!\!\!{\Gamma\left(-\delta_{A}+1\right)\over \pi 2^{2\delta_{A}}\Gamma\left(\delta_{A}\right)}
:e^{{\mu\over 2\pi}\ln{\varepsilon|x_{BC}|\over |x_{AB}||x_{AC}|}{\del^2\over \del\vec{x}_{AB}\cdot \del\vec{x}_{AC}}}:\\
&\hspace{-1.0cm}\times \prod_{(A,B,C)={\rm cycl}(1,2,3)}\int_{-\infty}^{+\infty} d^2k_A |k_A|^{2(\delta_{A}-1)} e^{i\vec{k}_A\cdot \vec{x}_{BC}
+{\mu\over \pi}|k_A|^2\ln(|x_{BC}|/\varepsilon)}\ .
\end{aligned}
\end{equation}
Here, the colons $:X:$ indicate that the differential operators are ordered to the rightmost, ensuring they do not act on any prefactors: schematically, this corresponds to the definition
$:e^{f(x)\partial^2_x}:\equiv \sum_{n=0}^{\infty}{1\over n!}f(x)^n\partial_x^{2n}$.
As in the two-point case~\eqref{2pt_all_orders_mG}, the momentum integrals can be evaluated by expanding the exponential in the final line. More efficiently, however, we can exploit the fact that the last two lines effectively encode a product of two-point correlators, allowing us to directly invoke the nonperturbative completions previewed in the previous section.
We thus arrive at an illuminating expression:
\begin{equation}
\begin{aligned}
\left\langle \prod_{A=1}^3{\cal O}_{\Delta_A}(x_A)\right\rangle^{\text{leading-log}}_{T\Tb}\!\!
&\!\!\!\!=C_{\Delta_1\Delta_2\Delta_3}\!\!\!\!\!\prod_{(A,B,C)={\rm cycl}(1,2,3)}
:e^{{\mu\over 2\pi}\ln{\varepsilon|x_{BC}|\over |x_{AB}||x_{AC}|}{\del^2\over \del\vec{x}_{AB}\cdot \del\vec{x}_{AC}}}:\\
&\times \prod_{(A,B,C)={\rm cycl}(1,2,3)}\langle {\cal O}_{\delta_{A}}(x_B){\cal O}_{\delta_{A}}(x_C)\rangle^{\text{leading-log}}_{T\Tb}\ .
\end{aligned}
\end{equation}
Here, $\delta_{C} = (\Delta_A + \Delta_B - \Delta_C)/2$ as before. Note that the exponential differential operator generates subleading logarithmic corrections in addition to the leading ones. Therefore, to remain conservative, we interpret only the leading logarithmic component of this expression as reliable. To the best of our knowledge, this represents a new result for $T\bar{T}$-deformed correlators.

%%%%%%%%%%%%%%%%%%%%%%%%%%%%%%%%%%%%%%%%%%%%%%%%%%%%%%%%%%%
\section{Discussion}\label{sec:discussion}

In this work, we have developed a systematic method for computing correlation functions in $T\bar{T}$-deformed conformal field theories using the massive gravity framework. Applying this approach, we computed the leading logarithmic corrections to two- and three-point functions at all orders in perturbation theory and obtained a new closed-form expression for the three-point case. Our results demonstrate how the massive gravity formulation provides a transparent and tractable framework for analyzing deformed correlators at finite coupling.

The results obtained in this work open several avenues for further exploration, both within the realm of $T\bar{T}$-deformed conformal field theories and in broader contexts where such deformations can be understood through the frameworks of quantum gravity and holography~\cite{Maldacena:1997re}.
A natural starting point is the effect of the deformation on the structure of conformal blocks, particularly in regimes such as large central charge, where the analytic behavior becomes more tractable. It remains an open question whether -- and in what way -- the deformation alters the asymptotic form or organization of these blocks, and what such changes might imply for the broader analytic structure of the theory. Even without a detailed mechanism, considerations of this kind may shed light on the interplay between irrelevant deformations and fundamental consistency conditions of conformal field theory, such as associativity and modularity.

These questions find a particularly rich setting in Liouville CFT, which -- with its continuous spectrum and well-controlled semiclassical limit -- serves as a natural testing ground. It would be of interest to examine how the $T\bar{T}$ deformation affects the dominant saddle-point geometries, and whether these admit a geometric interpretation within the massive gravity framework. Similarly, analyzing the impact of the deformation on Liouville correlators -- such as the DOZZ formula~\cite{Dorn:1994xn, Zamolodchikov:1995aa} -- may clarify how irrelevant deformations modify the exact analytic structure of the theory, including conformal block expansions, fusion rules, and crossing symmetry, without altering the underlying spectrum or operator content of the undeformed CFT. Given the central role of Liouville theory in the AGT correspondence~\cite{Alday:2009aq}, such deformations may also have implications for four-dimensional $\mathcal{N}=2$ gauge theories~\cite{Seiberg:1994rs} through the duality.

Beyond these structural considerations, $T\bar{T}$-deformed CFTs may also serve as useful toy models for Planckian physics~\cite{HR_NP_Planck}.  At distance scales shorter than the deformation scale, correlators exhibit qualitatively different behavior from their CFT counterparts: trans-Planckian oscillations give way to a regime in which correlations are typically suppressed by geometric randomness, in contrast to the power-law growth characteristic of undeformed theories. Moreover, the dependence of these correlators on distance becomes exponentially weaker, suggesting that the underlying geometric structure has been significantly washed out. This transition toward statistical incoherence at ultra-short scales may serve as a proxy for how locality and classical spacetime break down in a quantum gravitational regime, making the $T\bar{T}$ deformation a useful testing ground for probing the limits of field-theoretic resolution.

Finally, these insights naturally point toward a refined and constructive holographic perspective~\cite{HR_holography}. The massive gravity formulation offers a concrete framework for exploring holography with dynamical boundary gravity, extending the idea of~\cite{Hirano:2020nwq} to finite $T\bar{T}$ coupling. Rather than imposing a rigid radial cutoff by hand, as in~\cite{McGough:2016lol} (see also~\cite{Caputa:2020lpa} for a more closely related analysis relevant to our forthcoming work), fluctuations of boundary gravity may instead dynamically select a preferred bulk surface, offering a new perspective on how the $T\bar{T}$ deformation is encoded geometrically. Importantly, this framework naturally accommodates the inclusion of matter, making it possible to extend the holographic dictionary beyond the pure gravity sector. While our present analysis has focused on boundary correlators, the gravitational formulation invites further investigation into how features of bulk spacetime emerge from the deformed theory -- a direction we leave for future work.

%%%%%%%%%%%%%%%%%%%%%%%%%%%%%%%%%%%%%%%%%%%%%%%%%%
\appendix

\section{Details of perturbative calculations}
\label{app:details}

In Section~\ref{sec:perturbative}, we computed the first- and second-order $T\bar{T}$ corrections by expanding the undeformed CFT two-point function in a Taylor series. Here, we present this expansion explicitly up to quartic order in $\alpha$, which is required for the second-order calculation.

Since the full expression is lengthy, we separate the expansion into two parts. The terms up to quadratic order are given by
\begin{equation}
\begin{aligned}\label{A:quadratic}
{1\over |x+\alpha|^{2\Delta}}
&={1\over |x|^{2\Delta}}+\alpha^{\mu}\del_{\mu}{1\over |x+\alpha|^{2\Delta}}\biggr|_{\alpha=0}+{1\over 2}\alpha^{\mu}\alpha^{\nu}\del_{\mu}\del_{\nu}{1\over |x+\alpha|^{2\Delta}}\biggr|_{\alpha=0}+\cdots\\
&={1\over |x|^{2\Delta}}-{2\Delta x_{\mu}\alpha^{\mu}\over |x|^{2\Delta+2}}
-{\Delta\alpha^{\mu}\alpha^{\nu}\over |x|^{2\Delta+2}}\left(\delta_{\mu\nu}-{(2\Delta+2)x_{\nu}x_{\mu} \over |x|^2}\right)+\cdots\ ,
\end{aligned}
\end{equation}
while the cubic and quartic terms take the form
\begin{align}\label{A:quartic}
\hspace{-.5cm}
{1\over |x+\alpha|^{2\Delta}}&=\cdots +{1\over 3!}\alpha^{\mu}\alpha^{\nu}\alpha^{\rho}\del_{\mu}\del_{\nu}\del_{\rho}{1\over |x+\alpha|^{2\Delta}}\biggr|_{\alpha=0}
+{1\over 4!}\alpha^{\mu}\alpha^{\nu}\alpha^{\rho}\alpha^{\lambda}\del_{\mu}\del_{\nu}\del_{\rho}\del_{\lambda}{1\over |x+\alpha|^{2\Delta}}\biggr|_{\alpha=0}+\cdots\nn\\
&=+{\alpha^{\mu}\alpha^{\nu}\alpha^{\rho}\over 3!}{4\Delta(\Delta+1)\over |x|^{2\Delta+4}}
\biggl(\delta_{\mu\nu}x_{\rho}+\delta_{\mu\rho}x_{\nu}+\delta_{\nu\rho}x_{\mu}-{2(\Delta+2)x_{\mu}x_{\nu}x_{\rho}  \over |x|^2}\biggr)\nn\\
&\hspace{.5cm}+{4\Delta(\Delta+1)\alpha^{\mu}\alpha^{\nu}\alpha^{\rho}\alpha^{\lambda}\over 4!}{1\over |x|^{2\Delta+4}}
\biggl(\delta_{\mu\nu}\delta_{\rho\lambda}+\delta_{\mu\rho}\delta_{\nu\lambda}+\delta_{\nu\rho}\delta_{\mu\lambda}\\
&\hspace{.5cm}-{2(\Delta+2) \over |x|^2}\biggl(\delta_{\mu\lambda}x_{\nu}x_{\rho}+\delta_{\nu\lambda}x_{\mu}x_{\rho}+\delta_{\rho\lambda}x_{\mu}x_{\nu}  
+\delta_{\mu\nu}x_{\rho}x_{\lambda}+\delta_{\mu\rho}x_{\nu}x_{\lambda}+\delta_{\nu\rho}x_{\mu}x_{\lambda}  \biggr)\nn\\
&\hspace{.5cm}+{4(\Delta+2)(\Delta+3)x_{\mu}x_{\nu}x_{\rho}x_{\lambda}  \over |x|^4}\biggr)+\cdots\ .\nn
\end{align}
These expressions form the basis for the tensor structures in~\eqref{X_ij} and~\eqref{X_ijkl}.

To compute the second-order correction, we must next evaluate the combinatorics of the second term in~\eqref{2pt_second}. This proceeds in several steps:
(1) We begin by computing the tensorial factor $X_{\rm tensor}$ arising from the contraction of the Kronecker deltas produced by the functional derivatives with the tensor $\hat{X}_{ijkl}$. This yields
\begin{equation}
\begin{aligned}
X_{\rm tensor}&=\left(\delta^{ij}\delta^{kl}+\delta^{ik}\delta^{jl}+\delta^{il}\delta^{jk} \right)\hat{X}_{ijkl}\\
&={\Delta(\Delta+1)\over 3! |x_{12}|^{2\Delta+4}}\biggl((2\cdot 2+2+2)\times 3-{2(\Delta+2) \over |x_{12}|^2}(2+1+1)|x_{12}|^2\times 6\\
&\hspace{.45cm}+{4(\Delta+2)(\Delta+3)|x_{12}|^4  \over |x_{12}|^4}\times 3\biggr)
={\Delta(\Delta+1)\over 3! |x_{12}|^{2\Delta+4}}\times 12\Delta(\Delta+1)\ .
\end{aligned}
\end{equation}
This takes care of all the tensor contractions.
(2) Next, we compute the combinatorics associated with the functional derivatives acting on the Gaussian exponential. Since the tensorial structure has already been accounted for, we focus on one representative contraction structure among the three equivalent pairings: $\{(ij), (kl)\}, \{(ik), (jl)\}, \{(il), (jk)\}$. Fixing to the first structure, the fourth functional derivative
\begin{align}
{\delta^4\over\delta J^{i}\delta J^{j}\delta J^{k}\delta J^{l}}(J_{m}\Box^{-1}J^{m})(J_{m'}\Box^{-1}J^{m'})
\end{align}
acts such that each of the four derivatives hits a different source, producing a factor of 2 for each possible contraction within this pairing and a total of 4 ways to choose the sources. This gives a combinatorial factor
\begin{align}
X_{\rm derivative}=8\ .
\end{align}
(3) An additional factor arises from the position dependence of the sources. Recall that the operator structure in~\eqref{2pt_second} contains the combination
\begin{align}
\hspace{-.3cm}
\left(\!{\delta\over\delta J^{i}(x_1)}-{\delta\over\delta J^{i}(x_2)}\!\right)\!\left(\!{\delta\over\delta J^{j}(x_1)}-{\delta\over\delta J^{j}(x_2)}\!\right)\!
\left(\!{\delta\over\delta J^{k}(x_1)}-{\delta\over\delta J^{k}(x_2)}\!\right)\!\left(\!{\delta\over\delta J^{l}(x_1)}-{\delta\over\delta J^{l}(x_2)}\!\right).\!\!
\end{align}
For finite, non-singular contributions, each pair of derivatives in a given tensor structure must act at different points. Coinciding positions lead to divergences proportional to $\delta^2(0)$, which require renormalization (e.g., via point-splitting). Since each of the two derivative pairs has two valid position choices, we obtain
\begin{align}
X_{\rm position}=+4
\end{align}
(4) Finally, we include the overall numerical coefficient from expanding the exponential $e^{\mu j\Box^{-2}j}$, which yields $1+\mu j\Box^{-1}j+{\mu^2\over 2}(j\Box^{-1}j)^2+\cdots$. The fourth-order derivative thus picks out
\begin{align}
Y\equiv {\mu^2\over 2}\left({1\over 2\pi}\ln(|x-x'|/\varepsilon)\right)^2\ .
\end{align}
Putting all the pieces together, we obtain
\begin{align}
\mbox{2nd term of \eqref{2pt_second}}=X_{\rm tensor}X_{\rm derivative}X_{\rm position}Y={8\mu^2\Delta^2(\Delta+1)^2\over \pi^2}{\ln^2(|x_{12}|/\varepsilon)\over |x_{12}|^{2\Delta+4}}\ .
\end{align}
This result exactly matches the known expression~\eqref{2pt_all_orders}, confirming the validity of our prescription at second order.

%%%%%%%%%%%%%%%%%%%%%%%%%%%%%%%%%%%%%%%%%%%%%%%%%%%%%%%%%%

\section*{Acknowledgments}

SH would like to thank the department of mathematics at Nagoya University for their hospitalities during his visits where part of this work was done. The work of SH is supported in part by the National Natural Science Foundation of China under Grant No.12147219.

%%%%%%%%%%%%%%%%%%%%%%%%%%%%%%%%%%%%%%%%%%%%%%%%%%%%%%%%%%%%%


\begin{thebibliography}{40}

%\cite{Zamolodchikov:2004ce}
\bibitem{Zamolodchikov:2004ce} 
  A.~B.~Zamolodchikov,
  ``Expectation value of composite field $T$ anti-$T$ in two-dimensional quantum field theory,''
  hep-th/0401146.
  %%CITATION = HEP-TH/0401146;%%
  %119 citations counted in INSPIRE as of 11 Feb 2020
  
  %\cite{Smirnov:2016lqw}
\bibitem{Smirnov:2016lqw} 
  F.~A.~Smirnov and A.~B.~Zamolodchikov,
  ``On space of integrable quantum field theories,''
  Nucl.\ Phys.\ B {\bf 915}, 363 (2017)
  doi:10.1016/j.nuclphysb.2016.12.014
  [arXiv:1608.05499 [hep-th]].
  %%CITATION = doi:10.1016/j.nuclphysb.2016.12.014;%%
  %121 citations counted in INSPIRE as of 18 Nov 2019
  
 %\cite{Cavaglia:2016oda}
\bibitem{Cavaglia:2016oda} 
  A.~Cavagli\`a, S.~Negro, I.~M.~Sz\'ecs\'enyi and R.~Tateo,
  ``$T \bar{T}$-deformed 2D Quantum Field Theories,''
  JHEP {\bf 1610}, 112 (2016)
  doi:10.1007/JHEP10(2016)112
  [arXiv:1608.05534 [hep-th]].
  %%CITATION = doi:10.1007/JHEP10(2016)112;%%
  %107 citations counted in INSPIRE as of 21 Nov 2019 
  
   %\cite{Dubovsky:2017cnj}
\bibitem{Dubovsky:2017cnj} 
  S.~Dubovsky, V.~Gorbenko and M.~Mirbabayi,
  ``Asymptotic fragility, near AdS$_{2}$ holography and $ T\overline{T} $,''
  JHEP {\bf 1709}, 136 (2017)
  doi:10.1007/JHEP09(2017)136
  [arXiv:1706.06604 [hep-th]].
  %%CITATION = doi:10.1007/JHEP09(2017)136;%%
  %88 citations counted in INSPIRE as of 09 Mar 2020
  
      
 %\cite{Dubovsky:2018bmo}
 \bibitem{Dubovsky:2018bmo} 
   S.~Dubovsky, V.~Gorbenko and G.~Hern\'andez-Chifflet,
   ``$ T\overline{T} $ partition function from topological gravity,''
   JHEP {\bf 1809}, 158 (2018)
   doi:10.1007/JHEP09(2018)158
   [arXiv:1805.07386 [hep-th]].
   %%CITATION = doi:10.1007/JHEP09(2018)158;%%
   %70 citations counted in INSPIRE as of 09 Mar 2020
   
 %\cite{Tolley:2019nmm}
\bibitem{Tolley:2019nmm}
A.~J.~Tolley,
``$ T\overline{T} $ deformations, massive gravity and non-critical strings,''
JHEP \textbf{06}, 050 (2020)
doi:10.1007/JHEP06(2020)050
[arXiv:1911.06142 [hep-th]].
%64 citations counted in INSPIRE as of 02 Feb 2024  

%\cite{Tsolakidis:2024wut}
\bibitem{Tsolakidis:2024wut}
E.~Tsolakidis,
``Massive gravity generalization of $ T\overline{T} $ deformations,''
JHEP \textbf{09}, 167 (2024)
doi:10.1007/JHEP09(2024)167
[arXiv:2405.07967 [hep-th]].
%28 citations counted in INSPIRE as of 24 Jul 2025

%\cite{Hirano:2020nwq}
\bibitem{Hirano:2020nwq}
S.~Hirano and M.~Shigemori,
``Random boundary geometry and gravity dual of $ T\overline{T} $ deformation,''
JHEP \textbf{11}, 108 (2020)
doi:10.1007/JHEP11(2020)108
[arXiv:2003.06300 [hep-th]].
%34 citations counted in INSPIRE as of 10 Apr 2025


  %\cite{Cardy:2019qao}
\bibitem{Cardy:2019qao} 
  J.~Cardy,
  ``$T\overline T$ deformation of correlation functions,''
  arXiv:1907.03394 [hep-th].
  %%CITATION = ARXIV:1907.03394;%%
  %12 citations counted in INSPIRE as of 26 Nov 2019
  
  %\cite{Hirano:2024eab}
\bibitem{Hirano:2024eab}
S.~Hirano and M.~Shigemori,
``Conformal field theory on $ T\overline{T} $-deformed space and correlators from dynamical coordinate transformations,''
JHEP \textbf{07}, 190 (2024)
doi:10.1007/JHEP07(2024)190
[arXiv:2402.08278 [hep-th]].
%3 citations counted in INSPIRE as of 10 Apr 2025

%\cite{HR_NP_Planck}
\bibitem{HR_NP_Planck}
S.~Hirano and V.~Raj,
``Nonperturbative effects in $T\Tb$-deformed conformal field theories: A toy model for Planckian physics,''
[arXiv:2507.16262 [hep-th]].
 
  %\cite{Aharony:2023dod}
\bibitem{Aharony:2023dod}
O.~Aharony and N.~Barel,
``Correlation functions in $ \textrm{T}\overline{\textrm{T}} $-deformed Conformal Field Theories,''
JHEP \textbf{08}, 035 (2023)
doi:10.1007/JHEP08(2023)035
[arXiv:2304.14091 [hep-th]].
%25 citations counted in INSPIRE as of 15 Jul 2025
 
 
%\cite{Cui:2023jrb}
\bibitem{Cui:2023jrb}
W.~Cui, H.~Shu, W.~Song and J.~Wang,
``Correlation functions in the ${\text{TsT}}/T\overline{T }$ correspondence,''
JHEP \textbf{04}, 017 (2024)
doi:10.1007/JHEP04(2024)017
[arXiv:2304.04684 [hep-th]].
%21 citations counted in INSPIRE as of 15 Jul 2025

%\cite{Chen:2025jzb}
\bibitem{Chen:2025jzb}
L.~Chen, Z.~Du, K.~Liu and W.~Song,
``Symmetries and operators in $T\bar{T}$ deformed CFTs,''
[arXiv:2507.08588 [hep-th]].
%0 citations counted in INSPIRE as of 15 Jul 2025

 

%\cite{Barel:2024dgv}
\bibitem{Barel:2024dgv}
N.~Barel,
``Correlation functions in $ \textrm{T}\overline{\textrm{T}} $-deformed theories on the torus,''
JHEP \textbf{11}, 167 (2024)
doi:10.1007/JHEP11(2024)167
[arXiv:2407.15090 [hep-th]].
%5 citations counted in INSPIRE as of 18 Jul 2025


  %\cite{Conti:2018tca}
\bibitem{Conti:2018tca}
R.~Conti, S.~Negro and R.~Tateo,
``The $ \mathrm{T}\overline{\mathrm{T}} $ perturbation and its geometric interpretation,''
JHEP \textbf{02}, 085 (2019)
doi:10.1007/JHEP02(2019)085
[arXiv:1809.09593 [hep-th]].
%46 citations counted in INSPIRE as of 11 Nov 2020

%\cite{Cardy:2018sdv}
\bibitem{Cardy:2018sdv} 
  J.~Cardy,
  ``The $ T\overline{T} $ deformation of quantum field theory as random geometry,''
  JHEP {\bf 1810}, 186 (2018)
  doi:10.1007/JHEP10(2018)186
  [arXiv:1801.06895 [hep-th]].
  %%CITATION = doi:10.1007/JHEP10(2018)186;%%
  %71 citations counted in INSPIRE as of 09 Jan 2020  
    

%\cite{Hirano:2020ppu}
\bibitem{Hirano:2020ppu}
S.~Hirano, T.~Nakajima and M.~Shigemori,
``$ T\overline{T} $ Deformation of stress-tensor correlators from random geometry,''
JHEP \textbf{04}, 270 (2021)
doi:10.1007/JHEP04(2021)270
[arXiv:2012.03972 [hep-th]].
%27 citations counted in INSPIRE as of 10 Apr 2025

%\cite{He:2019vzf}
\bibitem{He:2019vzf}
S.~He and H.~Shu,
``Correlation functions, entanglement and chaos in the $ T\overline{T}/J\overline{T} $-deformed CFTs,''
JHEP \textbf{02}, 088 (2020)
doi:10.1007/JHEP02(2020)088
[arXiv:1907.12603 [hep-th]].
%83 citations counted in INSPIRE as of 24 Jul 2025

%\cite{He:2020qcs}
\bibitem{He:2020qcs}
S.~He,
``Note on higher-point correlation functions of the $T\bar T$ or $J\bar T$ deformed CFTs,''
Sci. China Phys. Mech. Astron. \textbf{64}, no.9, 291011 (2021)
doi:10.1007/s11433-021-1741-1
[arXiv:2012.06202 [hep-th]].
%37 citations counted in INSPIRE as of 24 Jul 2025

%\cite{He:2023kgq}
\bibitem{He:2023kgq}
S.~He, Y.~Sun and J.~Yin,
%``Systematic approach to correlators in TT{\textasciimacron} deformed CFTs,''
Phys. Rev. D \textbf{111}, no.8, 086016 (2025)
doi:10.1103/PhysRevD.111.086016
[arXiv:2310.20516 [hep-th]].
%14 citations counted in INSPIRE as of 24 Jul 2025

%\cite{He:2025ppz}
\bibitem{He:2025ppz}
S.~He, Y.~Li, H.~Ouyang and Y.~Sun,
``$T\overline{T}$ Deformation: Introduction and Some Recent Advances,''
[arXiv:2503.09997 [hep-th]].
%12 citations counted in INSPIRE as of 24 Jul 2025

  %\cite{Kraus:2018xrn}
\bibitem{Kraus:2018xrn} 
  P.~Kraus, J.~Liu and D.~Marolf,
  ``Cutoff AdS$_{3}$ versus the $ T\overline{T} $ deformation,''
  JHEP {\bf 1807}, 027 (2018)
  doi:10.1007/JHEP07(2018)027
  [arXiv:1801.02714 [hep-th]].
  %%CITATION = doi:10.1007/JHEP07(2018)027;%%
  %63 citations counted in INSPIRE as of 24 Nov 2019 


 
%\cite{Maldacena:1997re}
\bibitem{Maldacena:1997re}
  J.~M.~Maldacena,
  ``The large $N$ limit of superconformal field theories and supergravity,''
  Adv.\ Theor.\ Math.\ Phys.\  {\bf 2}, 231 (1998)
  [Int.\ J.\ Theor.\ Phys.\  {\bf 38}, 1113 (1999)]
  [arXiv:hep-th/9711200].
  %%CITATION = IJTPB,38,1113;%%
 

%\cite{Dorn:1994xn}
\bibitem{Dorn:1994xn}
H.~Dorn and H.~J.~Otto,
``Two and three point functions in Liouville theory,''
Nucl. Phys. B \textbf{429}, 375-388 (1994)
doi:10.1016/0550-3213(94)00352-1
[arXiv:hep-th/9403141 [hep-th]].
%512 citations counted in INSPIRE as of 11 Jul 2025

%\cite{Zamolodchikov:1995aa}
\bibitem{Zamolodchikov:1995aa}
A.~B.~Zamolodchikov and A.~B.~Zamolodchikov,
``Structure constants and conformal bootstrap in Liouville field theory,''
Nucl. Phys. B \textbf{477}, 577-605 (1996)
doi:10.1016/0550-3213(96)00351-3
[arXiv:hep-th/9506136 [hep-th]].
%858 citations counted in INSPIRE as of 11 Jul 2025

%\cite{Alday:2009aq}
\bibitem{Alday:2009aq}
L.~F.~Alday, D.~Gaiotto and Y.~Tachikawa,
``Liouville Correlation Functions from Four-dimensional Gauge Theories,''
Lett. Math. Phys. \textbf{91}, 167-197 (2010)
doi:10.1007/s11005-010-0369-5
[arXiv:0906.3219 [hep-th]].
%1433 citations counted in INSPIRE as of 11 Jul 2025

%\cite{Seiberg:1994rs}
\bibitem{Seiberg:1994rs}
N.~Seiberg and E.~Witten,
``Electric - magnetic duality, monopole condensation, and confinement in N=2 supersymmetric Yang-Mills theory,''
Nucl. Phys. B \textbf{426}, 19-52 (1994)
[erratum: Nucl. Phys. B \textbf{430}, 485-486 (1994)]
doi:10.1016/0550-3213(94)90124-4
[arXiv:hep-th/9407087 [hep-th]].
%3851 citations counted in INSPIRE as of 11 Jul 2025

 
 %\cite{HR_holography}
 \bibitem{HR_holography}
S.~Hirano and V.~Raj,
``$T\Tb$ braneworld holography,''
\textit{in preparation}.
   
   %\cite{McGough:2016lol}
\bibitem{McGough:2016lol} 
  L.~McGough, M.~Mezei and H.~Verlinde,
  ``Moving the CFT into the bulk with $T\bar T$,''
  arXiv:1611.03470 [hep-th].
  %%CITATION = ARXIV:1611.03470;%%
  %22 citations counted in INSPIRE as of 28 Mar 2018
     
    
 %\cite{Caputa:2020lpa}
\bibitem{Caputa:2020lpa}
P.~Caputa, S.~Datta, Y.~Jiang, Y.~Jiang,  and P.~Kraus,
``Geometrizing $ T\overline{T} $,''
JHEP \textbf{03}, 140 (2021)
[erratum: JHEP \textbf{09}, 110 (2022)]
doi:10.1007/JHEP03(2021)140
[arXiv:2011.04664 [hep-th]].
%31 citations counted in INSPIRE as of 24 Nov 2023


 
 
\end{thebibliography}
\end{document}